\documentclass[aps,prl,reprint,nobalancelastpage,nofootinbib]{revtex4-2}
\pdfoutput=1

\usepackage[utf8]{inputenc}

\usepackage{amsmath}
\usepackage{amssymb}
\usepackage[all]{xy}
\usepackage{tikz}
\usetikzlibrary{calc}

\usepackage[toc,page]{appendix}
\usepackage{enumerate}
\usepackage{tensor}
\usepackage{booktabs}
\usepackage{bbm}
\usepackage{youngtab}
\usepackage{slashed}
\usepackage{multirow}
\usepackage{makecell}
\usepackage{float}
\usepackage{comment}
\usepackage{verbatim}
\usepackage{xcolor}
\usepackage[framemethod=tikz]{mdframed}
\usepackage{soul}
\usepackage{mathtools}
\usepackage[inline]{enumitem}
\usepackage{alphabeta}
\usepackage{accents}

\usepackage{amsthm}

\newcommand{\der}{\partial}
\newcommand{\de}{\mathrm{d}}

\newcommand{\e}{\mathrm{e}}
\newcommand{\I}{\mathrm{i}}

\allowdisplaybreaks

\usepackage{hyperref}
\hypersetup{
    colorlinks=true,
    linkcolor=purple,
    citecolor=purple,
    filecolor=purple,      
    urlcolor=purple,
}

\begin{document}

\title{A mechanism for freezing moduli into Minkowski spacetime}
\author{Flavio Tonioni}
\email{flavio.tonioni@kuleuven.be}
\affiliation{Instituut voor Theoretische Fysica, KU Leuven, Celestijnenlaan 200D, B-3001 Leuven, Belgium}

\begin{abstract}
We discuss FLRW-cosmologies with negatively-curved spatial slices that induce a late-time freezing of all the moduli of an effective field theory through Hubble friction, independently of the moduli-space curvature. This holds for pure moduli, which appear ubiquitously in string compactifications. Crucially, the cosmological solutions approach a Milne universe. Hence, the mechanism we describe in fact corresponds to late-time moduli freezing into Minkowski spacetime.
\end{abstract}

\maketitle

\section{Introduction}
A universal feature of string compactifications is the presence of a large number of moduli.
Such moduli challenge phenomenology in that massless (or very light) fields disfavor epochs of accelerated cosmic expansion and may mediate unobserved long-range interactions.
Finding mechanisms to generate large masses for these fields is however a complex task.
Among others, the Dine-Seiberg problem represents a general argument as to why this is the case \cite{Dine:1985he}, relating to the fact that all string-theoretic potentials are generated dynamically, with perturbative expansions that also control the validity of the effective field theory (EFT) itself.
In fact, proposed mechanisms that are argued to allow for lower-dimensional vacua with Minkowski or de Sitter (dS) geometries require a delicate interplay of (non)perturbative effects such that the moduli are stabilized somewhere in between the bulk and the asymptotic regions \cite{Kachru:2003aw, Balasubramanian:2005zx}; for recent reviews, see refs.~\cite{Cicoli:2023opf, McAllister:2023vgy}.
Working in intermediate regions, however, confronts one with the issue of controlling all corrections: see ref.~\cite{McAllister:2024lnt} for a recent account.
Arguments based on the completion of an EFT into a full quantum-gravity framework, together with circumstantial evidence, actually led some to hypothesize the absence of potentials able to stabilize all moduli in asymptotic regions of moduli space of string compactifications as a general fact, at least if the potential is positive \cite{Obied:2018sgi, Ooguri:2018wrx}; for a review of the criticisms to dS constructions in string compactifications, see ref.~\cite{Danielsson:2018ztv}.

In this note, we describe a general mechanism by which Friedmann-Lemaître-Robertson-Walker (FLRW) cosmologies with negatively-curved spatial slices dynamically freeze all moduli, at late-enough times.
Here, by ``moduli'', we truly mean (pseudo)scalar fields with no potential.
The intuition behind the freezing mechanism is extremely simple.
On the one hand, the curvature term is a fluid that saturates the strong energy condition (SEC).
On the other hand, pure moduli just represent a kinating fluid, whose energy density decays more quickly than
that of a SEC-saturating fluid.
Therefore, given enough time, only the energy density of the curvature fluid survives.
Relative to the latter, the moduli kinetic energy must vanish.
It is in this sense that in the present work we intend the ``freezing''.
The dynamical mechanism is independent of the moduli-space curvature and of the details of the underlying string compactification.
A universe whose only component is its own negative spatial curvature, known as the Milne universe, is actually the interior of a future lightcone in Minkowski geometry, in suitable coordinates.
Hence, what we achieve is a dynamical process to freeze all moduli into Minkowski spacetime.
This mechanism works exactly in the absence of a potential, thus circumventing all arguments related to the large slope and/or negative convexity of positive scalar potentials in string compactifications.
The scope of a generic scenario where all moduli are frozen into a Minkowski geometry goes beyond describing the current universe.
It may serve as a dynamical explanation of cosmological initial conditions with small field kinetic energy, which helps modeling an expanding universe with a transient phase of acceleration, even in past epochs of the cosmological history.

The rest of this note is organized as follows.
In sec. \ref{sec.: dynamical moduli freezing}, we present the details of the curvature-induced freezing mechanism.
In sec. \ref{sec.: example: GKP compactifications}, we discuss a simple example.
Finally, a general discussion is in sec. \ref{sec.: discussion}.
Additional mathematical details are in app. \ref{app.: exact solution}.

\section{Dynamical moduli freezing} \label{sec.: dynamical moduli freezing}
Throughout this note, we study a set of $n$ moduli $\varphi^a$ that are associated to a moduli-space metric
\begin{equation} \label{field-space metric}
    \de s_{n}^2 = G_{ab} (\varphi) \, \de \varphi^a \de \varphi^b,
\end{equation}
with a corresponding Christoffel connection ${\Gamma^a}_{bc}$; Einstein summation is always implied on the field-space indices $a=1,\dots,n$.
Curved moduli spaces emerge ubiquitously in string compactifications and in supergravity constructions.
Let the non-compact $d$-dimensional spacetime be described by the FLRW-metric
\begin{equation} \label{FLRW-metric}
    d \tilde{s}_{1,d-1}^2 = - \de t^2 + a^2(t) \, \biggl[ \dfrac{\de r^2}{1 - k r^2} + r^2 \de \Omega_{d-2}^2 \biggr].
\end{equation}
Here, $a$ is the scale factor, which defines the Hubble parameter $\smash{H = \dot{a}/a}$, and $k$ is the constant curvature of the spatial slice.
Finally, we consider the presence of a set of homogeneous and isotropic fluids of energy densities $\rho^\alpha$ and pressures $p^\alpha = (q^\alpha-1) \rho^\alpha$, for $\alpha=1,\dots,b$.
A barotropic fluid has an equation of state $p = (q-1) \rho$, where $q$ is a constant in a multitude of cases including curvature anisotropies, radiation, matter, and a cosmological constant, with $q=2, d/(d-1), 1, 0$, respectively.

In this setup, the complete cosmological equations can be written as
\begin{subequations}
\begin{align}
    & \ddot{\varphi}^a + {\Gamma^a}_{bc} \dot{\varphi}^b \dot{\varphi}^c + (d-1) H \dot{\varphi}^a = 0, \label{moduli FRW-KG eq.} \\[2.5ex]
    & \dot{\rho}^\alpha + (d-1) H (p^\alpha + \rho^\alpha) = 0, \label{fluid continuity eq.} \\
    & H^2 = \dfrac{2 \kappa_d^2}{(d-1) (d-2)} \biggl[ \dfrac{1}{2} \, G_{ab} \dot{\varphi}^a \dot{\varphi}^b + \sum_{\alpha=1}^b \rho^\alpha \biggr] - \dfrac{k}{a^2}, \label{moduli-fluid Friedmann eq.}
\end{align}
\end{subequations}
where $\kappa_d$ is the gravitational coupling.
A combination of eqs.~(\ref{moduli FRW-KG eq.}, \ref{fluid continuity eq.}, \ref{moduli-fluid Friedmann eq.}) also gives the useful relationship
\begin{equation} \label{moduli-fluid acceleration eq.}
    \dot{H} = - \dfrac{\kappa_d^2}{d-2} \, \biggl[ G_{ab} \dot{\varphi}^a \dot{\varphi}^b + \sum_{\alpha=1}^b (p^\alpha + \rho^\alpha) \biggr] + \dfrac{k}{a^2}.
\end{equation}
The continuity eqs.~(\ref{fluid continuity eq.}) for constant values of $q^\alpha$ can be solved as $\smash{\rho^\alpha = \rho^\alpha_0 \, (a_0/a)^{(d-1) q^\alpha}}$, where $a_0 = a(t_0)$ is the scale factor at an initial time $t_0$.
Hence, the curvature term can be formally treated as an on-shell fluid with energy density $\rho_k = - [(d-1)(d-2)/(2 \kappa_d^2)] \, k/a^2$ and equation of state given by the parameter
\begin{equation} \label{FLRW-curvature q-parameter}
    q_k = 1 - \dfrac{d-3}{d-1}.
\end{equation}
Noticeably, this means that the negative-curvature fluid, i.e. the one with $k=-1/\ell^2$, for an arbitrary length (or time) scale $\ell$, saturates the SEC.

\subsection{Milne universe}
In the presence of a potential $V=V(\varphi)$ for the fields, eqs.~(\ref{moduli-fluid Friedmann eq.}, \ref{moduli-fluid acceleration eq.}) do not allow one to express the kinetic energy $\smash{T = G_{ab} \, \dot{\varphi}^a \dot{\varphi}^b/2}$ in terms of the Hubble parameter and of the fluid energy densities.
In the Friedmann eqs.~(\ref{moduli-fluid Friedmann eq.}, \ref{moduli-fluid acceleration eq.}), one generally finds the linear combinations $T+V$ and $T-V$, respectively.
However, if $V=0$, then $T$ is a variable that drops out of the equation governing the evolution of the scale factor.
In fact, a simple manipulation leads to the equation
\begin{equation} \label{master eq.}
    \dfrac{(d-1)(d-2)}{2 \kappa_d^2} H^2 + \dfrac{(d-2)}{2 \kappa_d^2} \dot{H} = \sum_\alpha \dfrac{2 - q^\alpha}{2} \rho^\alpha.
\end{equation}
The $a$-dependence of the fluid energy densities is known. Hence, the above is a second-order ordinary differential equation (ODE).
If the only fluid is negative FLRW-curvature, i.e. $k=-1/\ell^2$, the equation takes the form
\begin{equation} \label{fundamental eq.}
    H^2 + \dfrac{1}{d-1} \dot{H} = \dfrac{d-2}{d-1} \dfrac{1}{\ell^2 a^2}.
\end{equation}
One can find the analytic solution to this equation: details are in app. \ref{app.: exact solution}.
For the case where the only fluid is the negative FLRW-metric spatial curvature, a previous derivation of
the exact solution appeared in ref.~\cite{Bergshoeff:2008be}.

Here, we can shortly illustrate a way in which
the exact solution can be found.
Let us reparameterize the time $t$ in terms of a time coordinate $\sigma$ defined via \cite{Bergshoeff:2008be}
\begin{equation} \label{sigma-time}
    \de t^2 = \dfrac{\de \sigma^2}{\displaystyle \dfrac{\sigma^2}{\sigma_p^2} \, \biggl(\dfrac{\ell}{\sigma}\biggr)^{\!\!2(d-1)} + 1},
\end{equation}
for a real constant $\sigma_p$ that relates to the initial conditions.
Then, eq.~(\ref{master eq.}) is solved by
\begin{equation} \label{scale factor in sigma-time}
    a(\sigma) = \dfrac{\sigma}{\ell}.
\end{equation}
The fact that this is the only solution is discussed in more detail in app. \ref{app.: exact solution}.
In particular, we observe that at late-enough times it is possible to approximate $t(\sigma \sim \infty) \simeq \sigma$.

The main observation of the present note consists in the fact that, in terms of the cosmological time $t$, the late-time behavior of the scale factor is such that
\begin{equation} \label{milne-universe scale factor}
    a_\infty(t) = \dfrac{t}{\ell},
\end{equation}
which corresponds to a Milne universe.
Therefore, by plugging this solution into either eq.~(\ref{moduli-fluid Friedmann eq.}) or eq.~(\ref{moduli-fluid acceleration eq.}), one finds the requirement that
\begin{equation} \label{constant-field solutions}
    \varphi_\infty^a(t) = \varphi^a_\infty,
\end{equation}
where $\smash{\varphi^a_\infty \in \mathbb{R}}$ are constants.
One can immediately see that eq.~(\ref{constant-field solutions}) is a solution of eq.~(\ref{moduli FRW-KG eq.}).
For generic initial conditions, this solution is technically correct in the true asymptotic regime $t \sim \infty$, where the scale factor takes exactly the form in eq.~(\ref{milne-universe scale factor}).
One might fear that, before their kinetic energy dies off, the scalars might travel a long distance.
Actually, by an asymptotic expansion of the exact solution, one can verify that the scalar-field displacement falls off in a power-law form.
Hence, the field displacement is finite.
In string compactifications, this ensures that, if the EFT perturbative expansions are well-motivated at the initial time, then the dynamical evolution will not invalidate the expansions themselves at any time: for instance, a theory defined at large internal volume will always evolve within a large-volume regime. The field displacement can be computed analytically; again, we refer to app. \ref{app.: exact solution} for details.

We can interpret the vanishing of the moduli kinetic energy as being due to Hubble friction.
To see this, we can observe that pure moduli represent a perfectly stiff fluid, with a parameter
\begin{equation} \label{moduli q-parameter}
    q_\varphi = 2.
\end{equation}
It is indeed trivial to verify that, in the absence of a potential, the scalar field equation implies the continuity equation for a fluid with $\rho_\varphi = p_\varphi = G_{ab} \, \dot{\varphi}^a \dot{\varphi}^b/2$.
Comparing with eq.~(\ref{FLRW-curvature q-parameter}), the energy densities associated to the negative spatial FLRW-curvature and the moduli fall off with the scale factor as
\begin{subequations}
    \begin{align}
        \rho_k & = \rho_{k 0} \, \Bigl( \dfrac{a_0}{a} \Bigr)^{\!2}, \\
        \rho_\varphi & = \rho_{\varphi 0} \, \Bigl( \dfrac{a_0}{a} \Bigr)^{\!2(d-1)}. \label{moduli energy density}
    \end{align}
\end{subequations}
Hence, by a reasoning that is typical in cosmology, eventually the fluid with the lowest $q$-parameter dominates over the others.
One can also check that the scale factor in eq.~(\ref{milne-universe scale factor}) determines a late-time Hubble parameter $H_\infty(t) = 1/t$.
Hence, eq.~(\ref{moduli-fluid Friedmann eq.}) requires $\smash{\dot{\varphi}_\infty^a = 0}$.
We stress that none of these conclusions depends on the curvature of the moduli space.

If there are multiple fluids, the individual energy densities fall off more quickly for larger $q$-parameters.
Hence, after long enough, only the fluid with the smallest value of $q$ survives.
Therefore, the mechanism we presented keeps working even if we include any number of SEC-fulfilling fluids, such as radiation, matter and curvature anisotropies.

\subsection{Minkowski spacetime}
As is well known, the Milne universe coincides with the interior of a future lightcone in Minkowski spacetime \cite{Mukhanov:2005sc}.
To see this, we can define the radial coordinate
\begin{equation}
    \chi(r) = \mathrm{artanh} \, \biggl[ \dfrac{r}{\sqrt{r^2 + \ell^2}} \biggr],
\end{equation}
which is such that $\smash{\de \chi^2 = \de r^2 / (r^2 + \ell^2)}$.
With this coordinate choice, if the scale factor is $a(t)=t/\ell$, the metric in eq.~(\ref{FLRW-metric}) reads
\begin{equation} \label{milne metric}
    d \tilde{s}_{1,d-1}^2 = - \de t^2 + t^2 \Bigl[ \de \chi^2 + \mathrm{sinh}^2 \, \chi \, \de \Omega_{d-2}^2 \Bigr].
\end{equation}
Now, we can write the Minkowski metric as
\begin{equation} \label{minkowski metric}
    d s_{\mathbb{M}^{1,d-1}}^2 = - \de \tau^2 + \de \rho^2 + \rho^2 \, \de \Omega_{d-2}^2.
\end{equation}
By the coordinate transformation
\begin{subequations}
    \begin{align}
        \tau & = t \, \mathrm{cosh} \, \chi, \\
        \rho & = t \, \mathrm{sinh} \, \chi,
    \end{align}
\end{subequations}
we recover exactly the negatively-curved FLRW-metric with $a(t)=t/\ell$.
Crucially, the late-time regime $t \sim \infty$ in the Milne universe is a truly late-time regime in Minkowski time $\tau \sim \infty$, at any radial distance $\chi$ from the coordinate origin.

Solutions where the late-time scale factor is linear in cosmological time may be expected
on fairly general grounds
in FLRW-universes with negative spatial curvature, as the latter saturates the SEC.
However, such solutions are not necessarily
the Milne universe, and hence not Minkowski spacetime, after a coordinate redefinition.
To see this, let
\begin{equation}
    a(t) = \alpha \, \dfrac{t}{\ell},
\end{equation}
where $k=-1/\ell^2$. Then, the Ricci scalar for the metric in eq.~(\ref{FLRW-metric}) reads
\begin{equation}
    \tilde{R} = \dfrac{(d-1)(d-2)}{t^2} \, \biggl( 1 - \dfrac{1}{\alpha^2} \biggr),
\end{equation}
which is clearly non-zero for any value except $\alpha=1$.
It is this specific value that corresponds to a Milne universe.
Scalar fields tend to have a ``tracker behavior'': their $q$-parameter is not constant and in many known examples it tracks down the lowest $q$-value of other fluids in the theory, at late times \cite{Copeland:1997et}.
For instance, with spatial curvature $k=-1/\ell^2$, if $\gamma^2 \geq 4/(d-2)$, the theory of a single scalar $\phi$ with a potential $V = \Lambda \, \e^{- \kappa_d \gamma \phi}$ admits a solution $a(t) = (1/\sqrt{1 - 4/[(d-2) \gamma^2]}) \, t / \ell$ \cite{Marconnet:2022fmx, Andriot:2023wvg}.
It is only in the limit $\gamma \to \infty$ that one recovers a Milne universe, which indeed corresponds to $V \to 0$.
This can be understood by noticing that it is the scalar potential that induces the tracker behavior: in the absence of a potential, a set of scalars is instead a purely kinating fluid, with constant parameter $q_\varphi=2$.

\subsection{Cosmic deceleration}
One might wonder whether the late-time Minkowski geometry can be attained along a trajectory undergoing accelerated expansion, in the FLRW-formulation. Unfortunately, here we show that this is not the case.

Given an FLRW-metric as in eq.~(\ref{FLRW-metric}), cosmic acceleration is defined as the set of conditions $\dot{a}>0$ and $\ddot{a}>0$.
In particular, the condition on the second time derivative is equivalent to the condition that $\epsilon = -\dot{H}/H^2<1$.
In terms of the time coordinate $\sigma$ defined in eq.~(\ref{sigma-time}), we can express the $\epsilon$-parameter as
\begin{equation}
    \epsilon(t) = 1 + \dfrac{d-2}{1 + \Bigl( \dfrac{\sigma(t)}{\ell} \Bigr)^{\!2(d-2)}}.
\end{equation}
This means that we have $\epsilon(t)>1$ at all times $t$, and in particular that $\smash{\lim_{t \to \infty} \, \epsilon(t) = 1^+}$.
Hence, in the FLRW-formulation, the expansion is always decelerated.
This does not necessarily represent a problem for exploiting the freezing mechanism into string embeddings of cosmic inflation, since the moduli might only need to be stabilized or frozen after the inflationary phase.
On the other hand, it does not provide a viable way for getting a transient quasi-dS expansion at late times.

\section{Example: GKP compactifications} \label{sec.: example: GKP compactifications}
A protypical example of a string compactifcation down to a Minkowski geometry with a set of unstabilized moduli is the class of Giddings-Kachru-Polchinski (GKP) backgrounds \cite{Giddings:2001yu}.
These are type-IIB compactifications on a Calabi-Yau orientifold $\mathrm{CY}'{\!\!}_3$ with RR-3- and NSNS-3-form fluxes in which the K\"{a}hler sector happens to be flat in all directions.
Here we illustrate how the dynamical freezing of the latter can work.

As a compactification Ansatz, the 10d Einstein-frame line element $d\hat{s}_{1,9}^2 = \hat{g}_{MN} \, \de x^M \de x^N$ is taken to be
\begin{equation} \label{warped_metric}
    d\hat{s}_{1,9}^2 = \e^{2 A (y)} \, d\tilde{s}_{1,3}^2 (x) + \e^{-2 A (y)} \, d\breve{s}_6^2 (y),
\end{equation}
where the function $\e^{2A}$ is the so-called warp factor. The blocks $\tilde{g}_{\mu \nu} (x)$ and $\breve{g}_{mn}(y)$ are a 4d metric depending only on the 4d coordinates $x^\mu$ and the metric of a 6d Calabi-Yau orientifold depending on the 6d internal coordinates $y^m$, respectively. Due to Lorentz invariance, the background axio-dilaton $\tau$ varies only over the internal manifold.
Similarly, the complex 3-form flux $G_3 = F_3 - \tau H_3$ only has internal indices and the self-dual 5-form flux has the form $\smash{\tilde{F}_5 = (1+\hat{*}_{10}) \, \de \alpha \wedge \tilde{\mathrm{vol}}_{1,3}}$, for a real scalar $\alpha = \alpha(y)$.
Finally, the theory includes a set of localized sources such as D- and anti-D-branes as well as O- and anti-O-planes, compatibly with the Ansatz metric.

The 4d components of Einstein's equations and the 5-form flux field equation can be combined in an enlightening way, giving \cite{Baumann:2010sx}
\begin{equation} \label{warp-factor/alpha}
\begin{split}
    \breve{\nabla}^2 \Phi_- = \tilde{R}_4 & + \dfrac{(\Phi_+ + \Phi_-)^2}{4 \, \mathrm{Im} \, \tau} \, G{}_3^- \, \breve{\cdot} \, \overline{G}{}_3^- \\
    & + \dfrac{2}{\Phi_+ + \Phi_-} \, \der \Phi_- \breve{\cdot} \, \der \Phi_- + T_{4,6},
\end{split}
\end{equation}
where $\smash{G_3^- = [(\breve{*}_6 G_3) - \I G_3]/2}$, $\smash{\Phi_\pm = \e^{4 A} \pm \alpha}$, and $\smash{T_{4,6}}$ is a combination of the energy-momentum tensor of the localized sources, which vanishes e.g. for appropriate D3-brane/O3-plane configurations; we use the notation
$\smash{(A_q \breve{\cdot} B_q) \, \breve{\mathrm{vol}}_n = A_q \wedge \breve{*}_n B_q}$, on an $n$-dimensional space.
The left handside integrates to zero over a compact manifold, so the same must happen with the right-handside.
In a 4d Minkowski spacetime, with $\smash{\tilde{R}_4=0}$, each term in the right handside must integrate to zero, too, which implies the conditions
\begin{subequations}
\begin{align}
    \breve{*}_6 G_3 & = \I G_3, \label{G_3/imaginary_self-duality} \\
    \e^{4 A} & = \alpha. \label{warp_factor=alpha}
\end{align}
\end{subequations}
The latter can also solve the remaining field equations.
In terms of a 4d effective theory, such fluxes stabilize the axio-dilaton and the complex-structure moduli.
However, the K\"{a}hler moduli are purely flat directions.
For instance, the overall volume is controlled by the imaginary part of the chiral supermultiplet $\rho = \theta + \I \, \xi$, whose field-space metric block can be read off the K\"{a}hler potential $\kappa_4^2 K = -3 \, \mathrm{ln} \, [-\I (\rho - \overline{\rho})]$.
The mechanism we presented above freezes both fields $\theta$ and $\xi$ into finite values, while solving the 10d equations of motion.

\section{Discussion} \label{sec.: discussion}
In this note, we have described a generic framework in which all the moduli of a given EFT are dynamically frozen into constant finite values in a Minkowski geometry.
The process takes place independently of the moduli-space curvature and of any possible compactification details, such as the internal geometry.
Strictly speaking, the freezing mechanism takes place in the asymptotic future.
However, throughout the evolution, the kinetic energy of the fields becomes parametrically smaller and smaller compared to the energy set by the Hubble scale, in an associated formulation in terms of a FLRW-cosmology with negative spatial curvature.
The Hubble-sized energy is stored in the spatial FLRW-curvature.

It is of course not surprising that fluids (help) freeze the scalar kinetic energy, relatively to the Hubble scale.
For instance, this is an integral element in the recent studies of kination phases in early moduli-driven stringy cosmologies, where radiation terminates the kination epoch \cite{Conlon:2022pnx, Apers:2024ffe}.
Hubble damping generally helps overcome the overshoot problem \cite{Brustein:1992nk}.
In fact, such mechanisms have been exploited extensively in the literature: see e.g. refs.~\cite{Barreiro:1998aj, Huey:2000jx, Brustein:2004jp, Conlon:2008cj, Andriot:2024jsh}.
What is special about negative spatial FLRW-curvature, however, is the fact that asymptotically it gives a perfect Milne universe, which is actually the Minkowski spacetime.
Hence, not only does the scalar kinetic energy vanish, compared to the overall energy scale of the problem (the Hubble scale, in a cosmological setting), but it does so into a Minkowski vacuum (and not necessarily in a cosmological setting).

The setup for the result is that of classical field theory.
The final state that one finds is a Minkowski vacuum, with all the scalars fixed at a finite value.
However, such scalars have no Lagrangian mass.
Hence, one should expect that their fluctuations might mediate long-range forces, and possibly imply the time dependence of
fundamental constants; see e.g. ref.~\cite{Cicoli:2018kdo} for an account.
Looking at the problem from a different angle, the freezing mechanism that we have presented is generally unstable under the presence of scalar potentials.
At some point, as the Hubble energy asymptotes to zero, any scalar potentials (that perhaps one may have consistently neglected at higher energies) might become relevant, thus changing the dynamics of the problem.
Nonetheless, if such potentials are irrelevant at least initially, the dynamics that we described might at least serve as an explanation as to why the field kinetic energy is small, and thus also circumvent the discussion on long-range forces.
This might be of use in constructions that require some degree of fine tuning of the initial conditions to accommodate for realizations of dark energy, such as in refs.~\cite{Gomes:2023dat, Casas:2024xqy}.
We leave studies of other phenomenological questions such as the cosmological moduli problem \cite{Banks:1993en, deCarlos:1993wie} for future work.

Although with a number of caveats, the instance of a mechanism for dynamical moduli freezing into a Minkowski solution presented in this note is applicable to string compactifications in generality.
It also circumvents the problems that would be associated to the claim that 10d supergravity solutions compactified to 4d Minkowski always admit a 4d massless scalar \cite{Andriot:2022yyj}.
Recently, Minkowski vacua with no running moduli have also been obtained in non-geometric backgrounds that are mirror duals of rigid Calabi-Yau manifolds \cite{Rajaguru:2024emw, Becker:2024ayh}.

\begin{acknowledgments}
\subsection*{Acknowledgments}
I am grateful to Thomas Van Riet and Timm Wrase for inspirational conversations and for their precious feedback on the manuscript.
My work is supported by the FWO Odysseus grant GCD-D0989-G0F9516N.
I would also like to thank the Erwin Schrödinger International Institute for Mathematics and Physics in Vienna, Austria, and the organizers for hosting the thematic programme ``The Landscape vs. the Swampland'', for providing me with a space to develop my ideas and for financial support during my stay.
\end{acknowledgments}

\appendix

\section{Exact solution} \label{app.: exact solution}

A solution $a=a(t)$ to eq.~(\ref{fundamental eq.}) can be found analytically; the overall sign is arbitrary, but we focus on solutions with a positive scale factor $a>0$.
It is convenient to express it as the inverse of a function $b$, i.e.
\begin{equation} \label{exact solution}
    a(t) = b^{-1}(t),
\end{equation}
which reads
\begin{equation} \label{b-function}
    b(t) = \ell t \, {}_2 F_{1} \biggl[ \dfrac{1}{2}, - \dfrac{1}{2} \dfrac{1}{d-2}, \dfrac{1}{2} \dfrac{2d-5}{d-2}; \dfrac{r}{t^{2(d-2)}} \biggr],
\end{equation}
where $\smash{{}_2 F_{1}={}_2 F_{1}(a,b,c;z)}$ is the ordinary hypergeometric function and $r \in \mathbb{R}$ is an integration constant.
An initial time gauge has been chosen for brevity.
Although this expression is complicated, it admits an extremely simple late-time behavior.
For simplicity, let $\smash{r = s \, t_p^{2(d-2)}}$, with $s = \pm 1$ and $t_p>0$.
An asymptotic expansion then reveals that
\begin{align*}
    b(t) = \ell t \biggl[ 1 - \dfrac{1}{2} \dfrac{s}{2d - 5} \, \biggl( \dfrac{t_p}{t} \biggr)^{\!\!2(d-1)} + O\biggl( \dfrac{t_p}{t} \biggr)^{\!\!4(d-1)} \biggr].
\end{align*}
At leading order in $t_p/t$, we may approximate the function $b$ with the function $b_\infty(t) = \ell t$, which leads immediately to the late-time scale factor in eq.~(\ref{milne-universe scale factor}).
Working at next-to-leading order is harder, but the expression of $b(t)$ motivates the expansion
\begin{equation}
    a(t) = \dfrac{t}{\ell} \biggl[ 1 - q \, \biggl(\dfrac{t_p}{t}\biggr)^{\!\!m} + O\biggl(\dfrac{t_p}{t}\biggr)^{\!\!m_1} \biggr],
\end{equation}
for a constant $q \in \mathbb{R}$ that is yet undetermined and two positive constants $m_1>m>0$.
Plugging this back into eq.~(\ref{fundamental eq.}) indicates that one must fix $m=2(d-2)$.
Plugging this solution into eq.~(\ref{moduli-fluid Friedmann eq.}), one finds the late-time behavior
\begin{align*}
    T(t) = \dfrac{(d-2) q_d}{\kappa_d^2 t_p^2} \, \biggl(\dfrac{t_p}{t}\biggr)^{\!\!2(d-1)} \, \biggl[ 1 + O\biggl(\dfrac{t_p}{t}\biggr)^{\!\!m_2} \biggr],
\end{align*}
for some positive constant $m_2>0$, and where we defined $q_d = (d-1)(2d-5) \, q$ for convenience.
This means that one must have $q>0$; the same expression is confirmed by plugging the scale factor into eq.~(\ref{moduli-fluid acceleration eq.}).
One can also check that we may write $\smash{T = T_0 \, (a_0/a)^{2 (d-1)} \, \bigl[ 1 + O(a_0/a)^{m_0} \bigr]}$, for some constant $m_0>0$ and some reference values $a_0$ and $T_0$, in agreement with the fact that the moduli represent a stiff fluid (see eq.~(\ref{moduli energy density})).
For instance, in the case of a single canonical field, where $T = \dot{\varphi}^2/2$, one finds
\begin{equation}
    \varphi(t) = \varphi_\infty - \dfrac{\sqrt{2 q_d}}{\kappa_d \sqrt{d-2}} \, \biggl(\dfrac{t_p}{t}\biggr)^{\!\!d-2} \, \biggl[ 1 + O\biggl(\dfrac{t_p}{t}\biggr)^{\!\!m_3} \biggr],
\end{equation}
up to a sign, where $\varphi_\infty$ is an integration constant and $m_3>0$ is another positive constant. The integrated scalar field excursion from an initial time $t_0$ until infinity, defined as $\smash{\delta = \kappa_d \int_{t_0}^\infty \de t \; \varphi(t)}$, is
\begin{align*}
    \delta = \dfrac{t_0}{d-3} \, \dfrac{\sqrt{q_d}}{\sqrt{d-2}} \, \biggl(\dfrac{t_p}{t_0}\biggr)^{\!\!d-2} \biggl[ 1 + O\biggl(\dfrac{t_p}{t_0}\biggr)^{\!\!m_3} \biggr],
\end{align*}
if $d>3$. One has $\delta=\infty$ if $d=3$.

The solution in eqs.~(\ref{exact solution}, \ref{b-function}) was first derived in ref.~\cite{Bergshoeff:2008be}.
Here we revisit this derivation to provide further understanding of the results in the main text.
Let the spacetime metric be parameterized as
\begin{equation}
    d \tilde{s}_{1,d-1}^2 = - f^2(\sigma) \, \de \sigma^2 + a^2(\sigma) \, h_{i j}(x) \, \de x^i \de x^j,
\end{equation}
where $h_{ij} (x)$ is the metric over a maximally-symmetric $(d-1)$-dimensional space $\mathbb{X}^{d-1}$ with constant curvature $k$, for $i,j = 1, \dots, d-1$.
Let $\varphi^a = \varphi^a(\sigma)$ be a set of scalars.
In terms of the time variable $\omega$, defined through
\begin{equation}
    \dfrac{\de \omega}{\de \sigma} (\sigma) = \dfrac{f(\sigma)}{a^{d-1}(\sigma)},
\end{equation}
the action for the scalars takes the form
\begin{equation}
    S[\varphi] = \dfrac{1}{2} \, \int \de \omega \int_{\mathbb{X}^{d-1}} \de^{d-1} x \; G_{ab}(\varphi) \, \dfrac{\de \varphi^a}{\de \omega} \dfrac{\de \varphi^b}{\de \omega},
\end{equation}
which is formally independent of the metric.
Hence, the scalars have trivial dynamics and they just trace out geodesics in moduli space.
The Einstein equations for the function $a(\sigma)$ then read
\begin{equation}
    \biggl( \dfrac{\de a}{\de \sigma} \biggr)^2 = \biggl[ \dfrac{\kappa_d^2 \, p^2}{(d-1)(d-2)} \, \dfrac{1}{a^{2(d-2)}} - k \biggr] \, f^2,
\end{equation}
where $p^2 = G_{ab}(\varphi) \, (\de \varphi^a / \de \omega) \, (\de \varphi^b / \de \omega)$ is a constant.
This explains the disappearing of the scalar dynamics from the equations governing the metric evolution that we observe in eq.~(\ref{master eq.}).
A general solution takes the form
\begin{equation}
    a^2(\sigma) = \Bigl( \dfrac{\sigma}{\ell} \Bigr)^{\!2},
\end{equation}
where
\begin{equation}
    f^2(\sigma) = \dfrac{1}{\ell^2} \, \dfrac{1}{\displaystyle \dfrac{\kappa_d^2 \, p^2}{(d-1)(d-2)} \, \biggl(\dfrac{\ell}{\sigma}\biggr)^{2(d-2)} - k}.
\end{equation}
Let us reparameterize time as in eq.~(\ref{FLRW-metric}), namely through the definition $\de t = f(\sigma) \, \de \sigma$.
In case $k=-1/\ell^2$, we have $a(\sigma(t)) = a(t)$, with $a=a(t)$ in eqs.~(\ref{exact solution}, \ref{b-function}), confirming the result.
In terms of the time coordinate $\sigma$, it is already clear that the late-time behavior of the solution is a Milne universe.
Indeed, we see the behavior $\smash{f^2(\sigma \sim \infty) \simeq 1}$, which trivially reduces the metric to the form in eq.~(\ref{milne metric}).
The late-time behavior is independent of the initial conditions.

\bibliographystyle{apsrev4-1}
\bibliography{refs.bib}

\end{document}